\documentclass[a4paper,twocolumn,nofootinbib,superscriptaddress]{revtex4}
\usepackage{amsmath,epsfig}
\usepackage{amssymb}
\usepackage{slashbox}

\newcommand{\thelog}{\log}

\newcommand{\imax}{L}

\begin{document}

\clubpenalty=9999
\widowpenalty=9999

\title{Travelling waves and impact-parameter correlations}

\author{S. Munier}
\affiliation{Centre de physique th{\'e}orique, \'Ecole Polytechnique,
CNRS, Palaiseau, France}
\author{G. P. Salam}
\affiliation{LPTHE, UPMC Univ.\ Paris 6, 
  Universit\'e Paris Diderot (Paris 7),
  CNRS UMR 7589, Paris, France}
\author{G. Soyez}
\affiliation{Brookhaven National Laboratory, Building 510, Upton, NY 11973, USA}

\date{\today}

\begin{abstract}
It is usually assumed that the high-energy evolution of partons in QCD
remains local in coordinate space. 
In particular, fixed impact-parameter scattering is thought to
be in the universality class of one-dimensional
reaction-diffusion processes as if the evolutions at different points
in the transverse plane became uncorrelated through rapidity evolution.
We check this assumption
by numerically comparing a toy model with QCD-like impact-parameter 
dependence
to its exact counterpart with uniform evolution
in impact-parameter space.
We find quantitative differences, 
but which seem to amount to a mere rescaling of the strong coupling
constant. Since the rescaling factor 
does not show any strong $\alpha_s$-dependence,
we conclude that locality is well verified, up to subleading terms at small
$\alpha_s$.
\end{abstract}

\maketitle


\section{Introduction}

Recently, equations describing evolution towards high energy in
QCD, including saturation effects \cite{GLR,MQ,B,K1,JIMWLK,W}, have
been proven \cite{MP1} to belong to the universality class of the
Fisher-Kolmogorov-Petrovsky-Piscounov (FKPP) equation. It has also
been realised \cite{MSh} that additional contributions were to be
included in order to fully satisfy unitarity, in particular, in a way
consistent with boost-invariance. Though only some partial results
exist for a more complete set of evolution equations
\cite{IT,KL,HIMST}, the evolution of partonic states and their
scattering at high energy is conjectured \cite{IMM} to be a
process of the reaction-diffusion type. The QCD evolution equations
are thus thought to belong to the universality class of the stochastic
FKPP (sFKPP) equation \cite{VSP} which reads
\begin{multline}
\partial_t u(t,x)=\partial^2_{x} u(t,x)+u(t,x)-u^2(t,x)\\
+\epsilon\sqrt{2(u(t,x)-u^2(t,x))}\,\nu(t,x)
\label{sFKPP}
\end{multline}
where $\nu$ is a normal Gaussian noise uncorrelated in space and time;
$t$ corresponds to the rapidity variable in QCD (or the logarithm of the energy),
and $x$ is the logarithm of the transverse scale (momentum or distance)
relevant to characterise the partons in the considered process, while
$\epsilon$, the strength of the noise, is of the order of the strong
coupling constant $\alpha_s$;
$u$ can be thought of an event-by-event scattering amplitude,
whose average over events is the physical amplitude.
Such an equation is supposed to hold at a specific impact parameter.

A convenient picture for high-energy scattering is the colour dipole
model \cite{M1}, whose numerical implementation in terms of
Monte-Carlo code event generators \cite{S1,S_MC,MS,AGL} has been
particularly useful to investigate saturation issues in QCD.  The
picture is the following: a hadron may be represented as a set of
dipoles at the time of the interaction, which is constructed by
successive dipole splittings ({\em i.e.} emissions of gluons in the
large-$N_c$ approximation) and mergings (or any other nonlinear
process that would tame the growth of their number) whose
probabilities are computed from QCD vertices.  The set of dipoles
resulting from this process eventually interacts with a target by
exchanging gluons.

One condition for the one-dimensional\footnote{Through this paper,
  by {\em one-dimensional} we refer to models with a single spatial
  dimension ({\em e.g.} the dipole size), in addition to the evolution
  variable ({\em e.g.} time or rapidity).} sFKPP equation to be
relevant to high energy scattering is that the evolution remains local
in impact parameter: indeed, if this was not the case, then
correlations between different points in impact-parameter space would
invalidate one-dimensional equations like (\ref{sFKPP}) in QCD.  At
first sight, it is not at all obvious that this condition should
hold: 
the dipole evolution kernel is singular and allows for splittings
into areas that lie very far from each other in coordinate space.
This nonlocal behaviour may {\em a priori} cause migrations of the dipole
chains of successive splittings over large distances in the transverse
plane, which in turn may induce correlations between different impact
parameters.

In Ref.~\cite{IMM}, 
this independence of the different impact parameters
was assumed on the basis of a simple analytical estimate
of the mean number of dipoles produced collinearly.
Capturing the full complexity of the QCD evolution 
to achieve a more precise understanding
still seems out of reach of an analytical approach.
On the other hand, to our knowledge, all recent numerical studies of
high-energy scattering at saturation (including gluon-number
fluctuations) and its relation to the sFKPP equation (see
{\em e.g.} \cite{S,EGBM}) relied on the one-dimensional approximation.

Our goal in this paper is to test the validity of a one-dimensional
formulation such as Eq.~(\ref{sFKPP}) in a model that has
impact-parameter dependence and that we shall study numerically. We
could consider the full colour dipole model and use the
already-existing Monte-Carlo code in Ref.~\cite{S_MC}, but its non
trivial 2-dimensional impact-parameter structure makes it
far too complex to handle, especially since
we want to run the evolution over a time long enough to observe the
asymptotic properties.  Instead, we shall build a simpler model, with
discretised dipole sizes and a single dimension of impact parameter,
which reproduces
the main features of full QCD as far as the
singularities are concerned.

The structure of the paper is as follows.
Sec.~\ref{sec2} recalls the essential features of the colour
dipole model and its interpretation in a 
statistical mechanics language.
In Sec.~\ref{sec3} we construct a toy model which incorporates an
impact-parameter dependence, at variance with existing models of QCD
evolution. We also introduce a corresponding model without impact
parameter which would be fully equivalent to the first one in the
absence of correlations in impact parameter.
Finally, in Sec.~\ref{sec4}, we present the numerical results
on travelling waves and their impact-parameter correlations
obtained within this model and compare with the model without impact
parameter dependence.


\section{\label{sec2}Analytical preliminaries}

\subsection{Scattering in the dipole model}

In the colour dipole model \cite{M1}, high-energy evolution is viewed in the
following way. Two hadrons (which are asymptotically sets
of colourless $q\bar q$ dipoles, for the simplicity of the argument)
develop under rapidity evolution highly occupied Fock-states
which themselves may be seen as collections of colour dipoles.
The building up of the states of each hadron
is specified by providing the splitting rate of a
dipole whose endpoints have transverse coordinates $(x_{0},x_1)$ 
into two dipoles $(x_{0},x_2)$ and $(x_{1},x_2)$ as the result
of a gluon emission at position $x_2$.
It reads~\cite{M1}
\begin{equation}
\frac{dP}{d(\bar\alpha y)}(x_{01}\rightarrow x_{02},x_{12})
=\frac{x_{01}^2}{x_{02}^2 x_{12}^2}
\frac{d^2 x_{2}}{2\pi}.
\label{split00}
\end{equation}
This splitting process is actually supplemented by
a saturation mechanism, which limits the local density of dipoles to
about $1/\alpha_s^2$.
Unfortunately, it has not been formulated in the dipole model.
(For a study of the problems that one has to face to incorporate
saturation effects in this framework, see {\em e.g.} Ref.~\cite{IST}).
It is sometimes seen as a progressive slowing down of the dipole emission
rate once the local density is of the order of $1/\alpha_s^2$ 
\cite{JIMWLK,onedim}.
It could also be recombination of dipoles, which corresponds to
the reabsorption of gluons \cite{IST}, or some form of colour
reconnection (``swings'') among the constituent gluons of the dipoles
\cite{AGL}. 
But the very nature of the mechanism is not crucial
to our discussion, and when building our toy model, we
pick the simplest one to implement numerically.

At the time of the interaction, pairs of dipoles 
made of one dipole from the left-moving hadron and one dipole
from the right-moving hadron,
may exchange gluons,
provided that they are of similar sizes
and sit at the same impact parameter.
It is actually simpler to think of the process in the
rest frame of one of the hadrons: then, only the moving hadron
(the probe) undergoes evolution, while the other one (the target)
remains in its bare asymptotic state.

We need to clarify what is meant by two dipoles sitting at the same
impact parameter (up to a distance of the order of their size).
When a highly evolved hadron is probed by an elementary dipole of
size $x_{01}=x_0-x_1$, the scattering amplitude $T$ is related to the density
$n$ in the following way:
\begin{multline}
T(y,x_{01},{\scriptstyle\frac{x_0+x_1}{2}})=\frac{\pi^2\alpha_s^2}{2}
\times
\int \frac{d^2 z_0}{2\pi} \frac{d^2 z_1}{2\pi}\\
\times\thelog^2\frac{|x_0-z_1|^2|x_1-z_0|^2}{|x_0-z_0|^2|x_1-z_1|^2}
n(y,z_{01},{\scriptstyle\frac{z_0+z_1}{2}}),
\end{multline}
where the summation goes over the positions of the dipoles in the Fock state
of the evolved hadron. 
The first argument of $T$ is the rapidity, the second one the dipole size
and the last one the impact parameter.
This formula means that 
$T$ roughly counts the dipoles which have a 
size of the order of $|x_{01}|$
and sit in a region of impact parameter delimited by a circle of radius
$|x_{01}|$
around the position (or impact parameter) $(x_{0}+x_{1})/2$.

From the dipole splitting probability (\ref{split00}), we can infer
the evolution equation for $T$. The simplest version of this equation
is obtained when the target is a very large nucleus, that is, made of
a large (formally infinite) number of bare dipoles. One gets
\begin{multline}
\partial_{\bar\alpha y}T(y,x_{01},{\scriptstyle\frac{x_0+x_1}{2}})
=\int \frac{d^2 x_2}{2\pi}\frac{x_{01}^2}{x_{02}^2 x_{12}^2}
\bigg[T(y,x_{02},{\scriptstyle\frac{x_0+x_2}{2}})\\
+T(y,x_{12},{\scriptstyle\frac{x_1+x_2}{2}})
-T(y,x_{01},{\scriptstyle\frac{x_0+x_1}{2}})\\
-T(y,x_{02},{\scriptstyle\frac{x_0+x_2}{2}})
 T(y,x_{12},{\scriptstyle\frac{x_1+x_2}{2}})\bigg]
\label{bk}
\end{multline}
This is the well-known Balitsky-Kovchegov (BK) equation \cite{B,K1}.
Its linear part is the Balitsky-Fadin-Kuraev-Lipatov (BFKL) 
equation \cite{BFKL}, while its nonlinear part accounts for multiple
scattering on the dense target. 
For a general target, 
additional unitarity effects  
must be taken into account \cite{IMM,MSh}, and the corresponding
equation is not yet fully known, though some partial results covering
the main features are available \cite{IT,MSW,KL,HIMST}.
For example, at large-$N_c$ and in the two-gluon-exchange
approximation, it has been shown \cite{IT} that (\ref{bk}) has to be
supplemented by a non-local noise term taking into account the
gluon-number fluctuations when the target is dilute.

\subsection{Travelling waves}

Recent research has established a link between high energy QCD
evolution and reaction-diffusion processes.  Indeed, the
Balitsky-Kovchegov equation at fixed impact parameter, in the
diffusive approximation\footnote{The {\em diffusive approximation}
  corresponds to expanding the BFKL kernel to second order.}, was
shown to be identical to the FKPP equation \cite{MP1}.  Moreover, as
recalled in the Introduction, it is thought that full high-energy QCD
beyond the approximations assumed to establish the BK equation lies in
the universality class of reaction-diffusion processes \cite{IMM}.
The stochastic version of the FKPP equation (see Eq.~(\ref{sFKPP}))
then becomes relevant, with a noise term of the order of the strong
coupling constant $\alpha_s$. The interesting outcome of this
correspondence consists in universal results for the scattering
amplitudes that can be analytically computed in the asymptotic regime
of very small $\alpha_s$ and large rapidities though, for larger
values of the coupling, the same behaviours are observed numerically
\cite{EGBM,S,onedim}. Even if the precise form of the evolution equations are
not fully known, all the models presented so far have shown the same
asymptotic properties.

Generally speaking, equations of the form~(\ref{bk})
admit travelling wave solutions 
in the variable $\thelog(1/x_{01}^2)$ when the dependency upon the impact
parameter is ignored.
The main features of these waves only depend on the
Mellin moments $\chi(\gamma)$ 
of the splitting probability~(\ref{split00}).
These solutions are attractors in the sense that a large class of
initial conditions (actually all initial conditions 
that are physical for the considered
processes) converge to them at large rapidity.
The position $\rho_s(y)\equiv\thelog Q_s^2(y)$ of the wave defines the saturation
scale $Q_s(y)$. Thus, the saturation scale and its rapidity dependence
emerge naturally from the travelling-wave solutions. At large
rapidities, the speed of the wave $\partial_{\bar \alpha y}\rho_s(y)$ is 
\cite{GLR,GBMS,MT}
\begin{equation}
V=\frac{\chi(\gamma_c)}{\gamma_c}
\label{Vmf}
\end{equation}
where $\gamma_c$ satisfies
\begin{equation}
\chi^\prime(\gamma_c)=\frac{\chi(\gamma_c)}{\gamma_c}.
\end{equation}
The wave front 
decays exponentially in the small-$|x_{01}|$ region like
\begin{equation}
T\sim \exp\left(-\gamma_c\thelog\frac{1}{x_{01}^2Q_s^2(y)}\right).
\label{shape_front}
\end{equation}
Geometric scaling \cite{SGBK} is the physical phenomenon that corresponds
to the fact that the amplitude only depends on the
combination $x_{01}^2 Q_s^2(y)$ rather than on each variable
separately.

If stochasticity is taken into account,
then the properties of the solutions are modified.
The travelling waves become noisy, 
corresponding to the fact that each individual event is 
made of a discrete number of dipoles introducing fluctuations
important in the tail of the front where the dipole density is of
order 1 (or, $T\sim \alpha_s^2$).
The logarithm of the squared saturation scale 
acquires a variance that grows
linearly with rapidity.
The velocity of the fronts, that is to say, the rates of
growth of $\thelog Q_s^2(y)$ averaged over the realisations, 
reach an asymptotic value
smaller than that coming out of the BK equation by 
a shift 
proportional to $1/\thelog^2(1/\alpha_s^2)$.
The physical scattering amplitude is obtained by averaging over
events, and gets tilted \cite{BD} with respect to
what it would be if the BK equation held.
Geometric scaling is broken \cite{MSh} and replaced by diffusive 
scaling \cite{MSh,IMM,IT}:
\begin{equation}
\langle T\rangle =  
{\rm erfc}
\left[\frac{\thelog(x_{01}^2 Q_s^2(y))}
{\sqrt{\bar\alpha y/\thelog^3(1/\alpha_s^2)}}\right],
\end{equation}
which reflects the fact that the variance of the front
position scales like $\bar\alpha y/\thelog^3(1/\alpha_s^2)$.

It is important to note that the wave velocity gets closer to
the mean-field
result~(\ref{Vmf}) when $\alpha_s$ decreases, and not surprisingly,
the variance of the front position gets smaller in the same limit, which is the 
sign that fluctuations get milder.
This limit corresponds to large maximum local densities of dipoles,
for which statistical fluctuations are indeed expected to vanish,
and thus a mean-field approximation
may indeed be justified.

These results are believed to be rigorously true for one-dimensional
models of reaction-diffusion type. 
Their extension to QCD actually relies in particular
on the postulate
that the impact-parameter variable does not play a role
in equations such as~(\ref{bk}) and in their stochastic
extensions, for the latter should be ``local'' enough.
However, this fundamental assumption has never been checked,
and this is precisely what we intend to do numerically in this paper.
Before we turn to this task (Sec.~\ref{sec3}), 
let us recall the arguments in favour of the
decoupling of the travelling waves at different impact parameters.

\subsection{Picture of the evolution including impact parameter}

\begin{figure}
\begin{center}
\epsfig{file=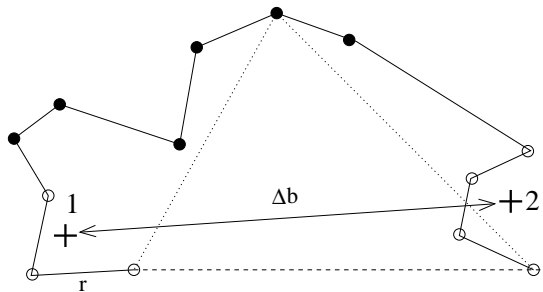,width=0.4\textwidth}
\end{center}
\caption{\label{scheme}Sample dipole configuration in impact-parameter space.
The points represent the gluons, and the line 
is the dipole chain after evolution. The 
initial dipole is represented by a dashed line, and the
first splitting by two dotted lines. For the 
two impact parameters denoted by a ``+'' sign (labelled 1 and 2)
their further respective evolutions
are believed to be uncorrelated.
The dipoles whose both endpoints are empty circles
are those which are seen at these peculiar impact parameters. For
clarity, we have omitted the large number of tiny dipoles produced at
every endpoints.
}
\end{figure}

Let us start with a single dipole at rest, and bring it gradually to a higher
rapidity. During this process, this dipole may be replaced by
two new dipoles, which themselves may split, and so on, eventually producing
a chain of dipoles.
Figure~\ref{scheme} pictures one realization of such a chain.

According to Eq.~(\ref{split00}), splittings to smaller-size dipoles
are favoured, and thus, one expects that the sizes of the dipoles
get smaller on the average,
and that in turn, the successive splittings become more local.
The dipoles around region ``1'' and those around region ``2'' should
have an independent evolution beyond the stage pictured in the figure: 
further splittings will not mix in impact parameter space, and thus,
the travelling waves around these regions should be uncorrelated.
For a dipole in region 1 of size $r$ to migrate to region 2, it should
first split into a dipole whose size is of the order of the
distance $\Delta b$ between regions 1 and 2, up to
a multiplicative uncertainty of order $1$. 
(We assume in this discussion 
that the dipoles in region 2 relevant to the propagation 
of the local travelling waves, that is, those
which are in the bulk of the wave front, also have sizes of order $r$).
Roughly speaking, the rate of such splittings may be estimated from
the dipole splitting probability~(\ref{split00}): 
it is of order $\bar\alpha (r^2/(\Delta b)^2)^2$,
while the rate of splittings of the same dipole into a dipole of similar
size in region 1 is of order $\bar\alpha$. Thus the first process is strongly suppressed
as soon as regions 1 and 2 are more distant than a few units of $r$.
Note that for $\Delta b \gtrsim 1/Q_s$, saturation may further reduce
the emission of the first, large, dipole leading to an even stronger
suppression of the estimated rate.

There is a second case that we should worry about.
What could also happen is that some larger dipole has, by chance, one of
its endpoints tuned to the vicinity of the coordinate
one is looking at (at a distance which is at most $|\Delta r|\ll 1/Q_s(Y)$),
and easily produces a large number of dipoles there.
In this case, the position of the travelling wave at that impact
parameter would suddenly jump.
If such events were frequent enough, then they would
modify the average wave velocity and thus the one-dimensional
picture. We may give a rough estimate of the rate at which
dipoles of size smaller than $\Delta r$ are produced.
%
Assuming local uniformity for the distribution of the emitting
dipoles, the rate (per unit of $\bar\alpha y$) of such events can be
written
\begin{equation}
  \int_{r_0 > \Delta r} \frac{d^2r_0}{r_0^2}
  \int_{\varepsilon < \Delta r} d^2\varepsilon\:
  n(r_0) \left(\frac{\varepsilon}{r_0}\right)^2
  \frac{1}{2\pi}\frac{r_0^2}{\varepsilon^2 (r_0-\varepsilon)^2},
\end{equation}
where we integrate over large dipoles of size $r_0 > \Delta r$
emitting smaller dipoles (of size $\varepsilon < \Delta r$) with a
probability $d^2\varepsilon\, r_0^2/(2\pi\varepsilon^2 (r_0-\varepsilon)^2)$. The factor
$(\varepsilon/r_0)^2$ accounts for the fact that one endpoint of the
dipole of size $r_0$ has to be in a given region of size $\varepsilon$ in
order to emit the dipoles at the right impact parameter.
To estimate this expression, we first use $n(r_0)=T(r_0)/\alpha_s^2$
and use for $T$ the simplified expression
\[
T(r_0) = \Theta(r_0 - 1/Q_s)\, + \,(r_0^2Q_s^2)^{\gamma_c}\, \Theta(1/Q_s - r_0),
\]
which splits the front into a saturated region ($r_0>1/Q_s$) and a
tail with geometric scaling ($r_0 < 1/Q_s$).
Using $r_0-\varepsilon \approx r_0$ in the emission kernel, the
integration is then easily performed and
one finds a rate whose dominant term is
\begin{equation}
\frac{\pi}{2\alpha_s^2}\frac{((\Delta r)^2Q_s^2)^{\gamma_c}}{1-\gamma_c}.
\end{equation}
For $(\Delta r)^2\ll(\alpha_s^2)^{1/\gamma_c}/Q_s^2$, that is,
ahead of the bulk of the front, this term is
parametrically less than 1 and is in fact of the order of the probability
to find an object in this region that contributes
to the normal evolution of the front \cite{BDMM}.
Hence there is no extra contribution due to the fact that there
are many dipoles around at different impact parameters.

However, it seems difficult to perform a precise calculation
of these effects beyond order-of-magnitude estimates, and
anyway,
this average picture may be spoiled by the statistical
event-by-event
fluctuations. So, we do not have any insight on how to estimate
more precisely this effect analytically. It is thus important
to perform a numerical simulation and check the correlations
of travelling waves at different impact parameters.


\section{\label{sec3}Construction of the model}

\subsection{\label{sec3A}Simplified model with impact-parameter dependence}

\subsubsection{Fock state evolution}

In order to arrive at a model that is tractable numerically, we only keep
one transverse dimension
instead of two in 3+1-dimensional QCD. 
However, we cannot consider genuine 2+1-dimensional QCD
because we
do not wish to give up the
logarithmic collinear singularities at $x_2=x_0$ and $x_2=x_1$.
Moreover, QCD with one dimension less has very different properties at high
energies \cite{IKLLSW}.
A splitting rate which complies with our requirements is:
\begin{equation}
\frac{dP}{d(\bar\alpha y)}=\frac{1}{4}\frac{|x_{01}|}{|x_{02}| |x_{12}|}dx_2.
\label{split0}
\end{equation}
We can further simplify this probability distribution
by keeping only its collinear and infrared asymptotics (as in \cite{CCS}).
If $|x_{02}|\ll |x_{01}|$ (or the symmetrical case $|x_{12}|\ll |x_{01}|$), 
the probability reduces to $dx_2/|x_{02}|$ ($dx_2/|x_{12}|$ resp.). 
The result of the splitting is a small dipole $(x_0,x_2)$ together with 
one close in size to the parent. So for simplicity we will just add the
small dipole to the system and leave the parent unchanged.
In the infrared region, a dipole of size $|x_{02}|\gg |x_{01}|$ 
is emitted with
a rate given by the large-$|x_{02}|$ limit of the above probability.
The probability laws~(\ref{split00}),(\ref{split0}) 
imply that a second dipole of
similar size should be produced while the parent dipole
disappears.
To retain a behaviour as close as possible to that in the collinear
limit, we will instead just generate a single large dipole and
maintain the parent. To do this consistently one must include a factor
of two in the infrared splitting rate, so as not to modify the average
rate of production of large dipoles.

In formulating our model precisely, let us focus first on the distribution of the 
sizes of the participating dipoles.
(The simplifying assumptions made above enable one to 
choose the sizes and the impact parameters of the dipoles successively).
We call $r$ the modulus of the emitted dipole, $r_0$ the modulus of its parent
and $Y=\bar\alpha y$.
The splitting rate~(\ref{split0}) 
reads, in this simplified model
\begin{equation}
\frac{dP_{r_0\to r}}{dY}
 = \theta(r-r_0) \frac{r_0 dr}{r^2} + \theta(r_0-r) \frac{dr}{r},
\end{equation}
and the original parent dipole is kept.
Logarithmic variables are the relevant ones here, so we introduce
\begin{equation}
\rho = \thelog_B(1/r) \qquad \text{or} \qquad r=B^{-\rho},
\end{equation}
where the base $B$ will later be set to 
two in the actual numerical calculations.
We can thus rewrite the dipole creation rate as
\begin{multline}
\frac{dP_{\rho_0\to \rho}}{dY}
 = \theta(\rho_0-\rho)\, B^{\rho-\rho_0}\, \thelog B\, d\rho\\
 + \theta(\rho-\rho_0)\, \thelog B\, d\rho.
\end{multline}
To further simplify the model,
we discretise the dipole sizes and
consider a lattice in $\rho$ with lattice spacing
$\Delta$ (which will later be set to one). 
This amounts to restricting the dipole sizes to negative
integer powers of $B^\Delta$.
The probability that a dipole at lattice site $i$ ({\em i.e.} a dipole
of size $B^{-i\Delta}$) creates a new
dipole at lattice site $j$ is
\begin{eqnarray}
\frac{dP_{i\to j}}{dY}
 &=& \int_{\rho_j}^{\rho_{j+1}} \frac{dP_{\rho_i\to\rho}}{dY}\\
 &=& \begin{cases}
\Delta \thelog B & j\ge i\\
(B^\Delta-1) B^{(j-i)\Delta} & j<i
\end{cases}.
\label{splitf}
\end{eqnarray}
The rates $dP_{i\pm}/dY$ for a dipole at lattice site $i$ to split to any
lattice site $j\ge i$ or $j<i$ respectively are then given by
\begin{eqnarray}
\frac{dP_{i+}}{dY} & = & \sum_{j=i}^{\imax-1} \frac{dP_{i\to j}}{dY}
 = \Delta \thelog B (\imax-i),\\
\frac{dP_{i-}}{dY} & = & \sum_{j=0}^{i-1} \frac{dP_{i\to j}}{dY}
 = 1-B^{-i\Delta},
\label{lifetimef}
\end{eqnarray}
where we have restricted the lattice to $0\le i <\imax$, for obvious
reasons related to the numerical implementation.

Now we have to address the question of the impact parameter of the emitted dipole.
In QCD, the collinear dipoles are produced near the endpoints of the 
parent dipoles. Let us take a parent of size $r_0$ at impact parameter $b_0$.
We set
the emitted dipole (size $r$) at the impact parameter $b$ such that
\begin{equation}
b=b_0\pm \frac{r_0\pm r\times s}{2}
\label{bf}
\end{equation}
where $s$ has uniform probability between 0 and 1. It is introduced to
obtain a continuous distribution of the impact parameter unaffected by
the discretisation of $r$.
This prescription is quite arbitrary in its details, but the latter
do not influence significantly the physical observables.
Each of the two signs that appear in the above expression
is chosen to be either $+$ or $-$ with equal weights.
We apply the same prescription 
when the emitted dipole is larger than its parent.

\subsubsection{Scattering amplitude}

We explained before that in QCD, the scattering amplitude 
of an elementary probe dipole
of size $r_i=B^{-i\Delta}$
with a dipole in an evolved Fock state
is proportional to the number of objects which have a size of the same
order of magnitude and which sit in an area of radius of order $r_i$ around the
impact point of the probe dipole.
Since in our case, the sizes are discrete, the amplitude is just given,
up to a factor, by the
number of dipoles that are exactly in the same bin of
size as the probe, namely
\begin{multline}
T(i,b_0)=(\alpha_s^2 / \Delta)
\times\# \{\text{dipoles of size $B^{-i\Delta}$} \\\text{at impact parameter $b$
such that }|b-b_0|<r_i/2\}.
\label{Tf}
\end{multline}

\subsubsection{Saturation}

We now have to enforce unitarity, that is the condition
\begin{equation}
T(i,b)\leq 1
\label{unitarity}
\end{equation}
for any $i$ and $b$.
This condition is expected to hold due to gluon saturation in QCD.
However,
saturation is not included in the
original dipole model.
Nevertheless, as argued in Section \ref{sec2}, the asymptotic
properties are not affected by the details on how we implement the
condition \eqref{unitarity}.
The simplest choice is to veto splittings that would locally 
drive the amplitude
to values larger than 1.
In practice, for each splitting that gives birth to a new dipole
of size $i$ at impact parameter $b$, we compute
$T(i,b)$ and $T(i,b\pm r_i/2)$, and throw away the
produced dipole whenever
one of these numbers gets larger than one.

Given the definition of the amplitude $T$, this saturation rule implies
that there is a maximum number of objects in each bin of size
and at each impact parameter, which is equal to $N_\text{sat}=\Delta/\alpha_s^2$.

\subsubsection{Note on the implementation}

The model is now completely specified by Eqs.~(\ref{splitf}),~(\ref{bf}),~(\ref{Tf})
and~(\ref{unitarity}).

The implementation of the dipole splittings is quite
straightforward, since the distribution of the sizes 
and impact parameters of the
produced dipoles are very simple.
At a given point along the rapidity evolution,
we choose the dipole that is going to split next.
The dipole lifetimes only depend on their sizes, 
and are simply given by the inverse rates $d(P_{i+}+P_{i-})/dY$ 
in Eq.~(\ref{lifetimef}).
Once this dipole has been chosen, we determine
its size and impact parameter according to Eqs.~(\ref{splitf}),~(\ref{bf}).
Next, we check that the conditions for the amplitude not to
violate unitarity are satisfied. If this is the case, the new dipole
is integrated in the Fock state. Otherwise, the dipole
is rejected.

The structure of the data has to be chosen carefully
to save CPU time.
Indeed, in order to compute the cross section $T$,
one needs to search for all dipoles of a given size
within a specified range of impact parameter about the probe dipole.
Given that the number of dipoles grows very quickly with rapidity (a priori
exponentially),
it is crucial to be able to search these dipoles in a time that does not
grow faster than the logarithm of their number.
To meet this requirement,
an array indexed by the discrete logarithmic size $i$ of the dipoles
points to binary trees of dipoles of identical sizes, ordered by their
impact parameter.

The number of objects grows very fast
due to the large number of tiny dipoles that may easily
be produced, and consequently,
the computer memory would be saturated quite early in the evolution.
Imposing the above-mentioned cutoff $\imax$ on the logarithmic size 
is not efficient enough to limit the number of objects.
The reason is that $\imax$ has to be taken quite high (we chose 50): 
if it is lower than
that, the evolution gets hampered within the range of rapidity that
is of interest for investigating saturation.
Hence we decided to enforce a lower cutoff 
also on the ratio $\kappa=B^{-i}/|b-b_N|$ of the size
of each produced dipole to the distance between its impact parameter $b$ and
any impact parameter $b_N$ at which we would measure cross sections.
Imposing this cutoff, we anticipate on the fact that a small dipole
very far from the impact parameters of interest
has only a tiny probability to migrate back through its 
further splittings;
$\kappa$ has of course to be taken sufficiently small, and varied in order 
to evaluate its effect.

Restrictions due to machine accuracy must also be addressed.
Indeed, when dipoles become small through evolution, one 
must be able to resolve numerically 
equally small distances in impact
parameter.
Since absolute impact parameters will be of the order of 1 
and are coded on 53 significant bits 
(typical double precision floating point numbers), 
we have to limit
the dipole sizes to about $i<53 \thelog B/\thelog 2$.
A way to overcome this limitation is to use arbitrary precision
arithmetic libraries \cite{gmp}, but this is both more memory and CPU costly so
we tried to avoid this solution.


\subsection{Fixed impact-parameter version}

We may now consider a similar model, but in which
there is no impact-parameter dependence.
We will call it ``FIP'' for Fixed Impact Parameter,
while the complete model will be termed ``AIP'' (Any Impact Parameter).

\subsubsection{Formulation}

One may understand how to ``remove'' the impact parameter from the AIP
model as follows. The essential point is that the interaction between
the probe and target is local in impact parameter. That means that if
a probe of size $r$ is to interact with a dipole of size $r$ in the
target, then the probe should also be roughly within a transverse
distance $r$ of the target dipole. When explicit impact parameter
information is maintained, one simply enforces this in one's
determination of probe-target interactions. If instead one discards
the impact parameter information, then one must find another way of
accounting for the fact that most target dipoles of size $r$ will be
too far from the probe to have a significant interaction with it.
The simplest is to observe that the probability of a target dipole
being close enough to the probe is $r/r_0$, where $r_0$ is the initial
size of the system. Therefore the number of dipoles of size $r$
sufficiently close to a specific fixed impact parameter, $n^{(f)}(r)$, is
given in terms of the total number of dipoles of size $r$, $n(r)$, by
the relation $n^{(f)}(r) = r/r_0\, n(r)$. One may reinterpret this in terms
of an FIP-specific branching probability, which relates to the normal
branching probability via
\begin{eqnarray}
\frac{dP_{i\to j}^{(f)}}{dY}
& = & \frac{r_j}{r_i} \frac{dP_{i\to j}}{dY}
 = B^{(i-j)\Delta} \frac{dP_{i\to j}}{dY}\\
& = & \begin{cases}
   \Delta \thelog B \,B^{(i-j)\Delta} & j\ge i\\
   (B^\Delta-1)                  & j <  i
   \end{cases}.
\end{eqnarray}
It is straightforward to verify that this reproduces the condition 
$n^{(f)}(r) = r/r_0\, n(r)$, and from eqs. (22,23) one then derives
\begin{eqnarray}
\frac{dP^{(f)}_{i+}}{dY} & = & \sum_{j=i}^{\imax-1} \frac{dP^{(f)}_{i\to j}}{dY} = 
\Delta \thelog B\frac{1-B^{-(\imax-i)\Delta}}{1-B^{-\Delta}},\\
\frac{dP^{(f)}_{i-}}{dY} & = & \sum_{j=0}^{i-1} \frac{dP^{(f)}_{i\to j}}{dY} = 
(B^{i\Delta}-1)\,i.
\end{eqnarray}

Of course, the saturation rule in the FIP model is that a dipole is not
produced if the number of objects of the same size is already larger than
$N_\text{sat}$.

%


\subsubsection{Numerical implementation}

The FIP model is quite straightforward to implement.
The splitting probabilities are as simple as those in the AIP model.
Since again the dipoles have discrete sizes,
it is enough to update an array indexed by the logarithmic dipole
size $i$, each cell of which
contains the corresponding number of dipoles.


\subsection{Expected properties of the amplitudes in these models}

At this stage, we can derive the properties of the travelling waves.
The first step is to write down the BFKL evolution of the dipole
densities. If $n_i$ is the number of dipoles at lattice site $i$, we have
\begin{equation}
\begin{split}
\partial_Y n_i 
 & =  \sum_j \frac{dP_{j\to i}^{(f)}}{dY} n_j \\
 & =  \sum_{j\ge i} \Delta \thelog B\,B^{(j-i)\Delta} n_j
     + \sum_{j<i} (B^\Delta-1) n_j.
\end{split}
\end{equation}
The amplitude $T$ is just $n_i/N_\text{sat}$.

The equation that corresponds to the BK equation, which preserves
unitarity, is obtained from the latter 
by taking the minimum of $n$ as given by the linear
evolution and $N_\text{sat}$,
at each rapidity step and for each dipole size.

The FIP model has the same number of variables as the FKPP equation:
one evolution variable and one spatial variable (which is the dipole
size) in which diffusion can occur.
%
It is a branching diffusion process in the discretised space
of dipole sizes, with
a saturation condition that limits the number of objects 
(dipoles) of each size.
From a general analysis of such models, we know that the realisations of
the evolution of the scattering amplitude are noisy travelling waves
whose properties may be obtained from the branching diffusion kernel.

The eigenfunctions of the kernel of the fixed-impact parameter evolution
are of the form $B^{-i\gamma\Delta}$ with 
the corresponding eigenvalues 
\begin{equation}
\begin{split}
\chi(\gamma) 
 & =  \sum_{k\le 0} \Delta \thelog B\,B^{-k\Delta}\,B^{k\gamma\Delta}
     + \sum_{k<0} (B^\Delta-1)\,B^{k\gamma\Delta}\\
 & =  \frac{\Delta \thelog B}{1-B^{(\gamma-1)\Delta}}
     + \frac{B^\Delta-1}{B^{\gamma\Delta}-1}.
     \label{eq:FIPkernel-gamma}
\end{split}
\end{equation}
Note that in the continuous limit obtained by letting $\Delta$ go to
0, one recovers the standard collinear approximation to the BFKL kernel:
\begin{equation}
\chi(\gamma)=\frac{1}{\gamma}+\frac{1}{1-\gamma},
\end{equation}
which confirms that we have kept the most important singularities.
Furthermore, one may verify that the $1/\gamma$ and $1/(1-\gamma)$
singularities are present independently of the value of $\Delta$.
From the expression Eq.~(\ref{eq:FIPkernel-gamma}) for the kernel, it
is possible to find the 
critical parameters that control the travelling waves for the
mean-field limit ({\em i.e.} the infinite $N_\text{sat}$ limit)
of the evolution with saturation. A numerical
solution of the condition $\gamma \chi'(\gamma)=\chi(\gamma)$ (for
$\Delta=1$ and $B=2$) leads to
\begin{equation}
\begin{split}
& \gamma_c = 0.69216\cdots,\qquad
\chi(\gamma_c) = 5.2314\cdots\\
& \text{and}\qquad
\chi'(\gamma_c) = 7.5582\cdots.
\end{split}
\end{equation}
The travelling waves proceed to larger values of $i$ 
as rapidity is increased. They
are characterised by their position $\rho_\text{s}$,
related to the saturation scale in QCD, and by the shape of their leading 
edge. 
Note that there is a certain freedom in the definition of the position
of the front: we will specify our choice in Sec.~\ref{sec4a}.
The large-rapidity velocity of the wave reads
\begin{equation}
\frac{d\rho_\text{s}}{dY}=\frac{1}{\thelog B}\chi'(\gamma_c),
\end{equation}
while the shape is given by
\begin{equation}
T(i)\sim B^{-\gamma_c (\Delta i- \rho_{\rm s})}
\end{equation}
in the forward part of the wave front defined by $i\Delta \gg \rho_\text{s}$.
For a finite $N_\text{sat}$, the velocity is in general
lower, and 
differs from the the mean-field velocity by amount of order
$1/\thelog^2 N_\text{sat}$.
Furthermore, the front position
acquires a dispersion from event to event,
which goes like $Y/\thelog^3 N_\text{sat}$ asymptotically
for large $Y$ and $N_\text{sat}$.

If the property that all impact parameters evolve independently
of each other is true,
travelling waves with the same properties are expected for
$T(i,b)$ in the AIP model, at each value of the impact parameter.
This is what we shall investigate numerically in the last section.


\section{\label{sec4}Numerical results}

We take as an initial condition a number $N_\text{sat}$ 
of dipoles of size 1 ($i=0$),
uniformly distributed in impact parameter between
$-r_0/2$ and $r_0/2$ in the case of the AIP model.
(This means that the impact parameters of the initial $N_\text{sat}$
dipoles are randomly chosen according to a flat
distribution\footnote{We have checked that starting with a fixed initial distribution of
  dipoles in impact parameter leads to the same asymptotic
  results. See {\em e.g.} Fig. \ref{plot_sigma2} further in this
  Section where the simulation for $N_\text{sat}=100$ up to $Y=4$ has
  been obtained by taking $N_\text{sat}$ dipoles at $b=0$ as an
  initial condition: only differences in the pre-asymptotic regime are
  observed.}).
We vary $N_\text{sat}$ between 10 and 200, and the cutoff $\kappa$
between $10^{-1}$ and $10^{-4}$.
The impact parameters $b_j$ that are considered are respectively
$0$, $10^{-6}$, $10^{-4}$, $10^{-2}$ and $10^{-1}$.
We set $\Delta=1$ and $B=2$ in all that follows.

The number of events generated is typically $10^4$ for each set of parameters, 
which allows one to measure the average and variance of the position 
of the travelling waves to a sufficient
accuracy.

The numerical data presented below were obtained using two independent
implementations of the Monte-Carlo event generator, which gave fully
consistent results.

\subsection{\label{sec4a}Amplitudes at a given impact parameter}

For the needs of the discussion in this section, convenient values 
of $N_\text{sat}$ and $\kappa$ are 25 and $10^{-2}$ respectively. 
The latter will be the default values, unless stated explicitly.

\begin{figure}
\begin{center}
\epsfig{file=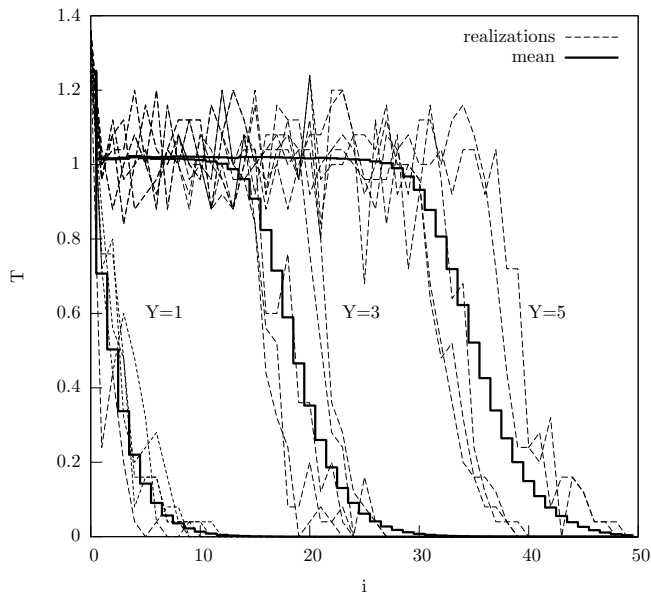,width=0.5\textwidth}
\end{center}
\caption{\label{plot_front}
Shape of the travelling-wave front in the AIP model at central
impact parameter, for three 
rapidities ($Y=1$, $3$ and $5$). 
Five different events are shown in dashed lines,
while the average over many events is displayed with 
continuous lines. $N_\text{sat}=25$ and $\kappa=10^{-2}$.
}
\end{figure}

First, we can observe very clearly the propagation of the travelling waves, in the
FIP model (not pictured)
as well as in the AIP model at different impact parameters 
(see Fig.~\ref{plot_front}; similar curves would be obtained in the
FIP model). Note that the different realisations shown in
Fig.~\ref{plot_front} sometimes overshoot 1. This is due to the fact
that the unitarity constraint is only imposed at a finite number of
points (in practice only 3, in order to save CPU time).
Adding more points reduces the excess but does not change the
final results for the wave front velocity and dispersion. 

To explore quantitatively the properties of the waves,
we define their position $\rho_\text{s}$ as the largest $i$
({\em i.e.} the smallest dipole size)
for which 
$T\geq 1/2$,
namely
\begin{equation}
\begin{split}
&N_\text{sat}\times T(\rho_\text{s},b_j)\geq \frac{ N_\text{sat}}{2}\\
\text{and}\qquad
&N_\text{sat}\times T(\rho_\text{s}+1,b_j)< \frac{ N_\text{sat}}{2}.
\end{split}
\end{equation}
In this section, we will be interested in measuring the 
$Y$-slope of the average of $\rho_\text{s}$ over events, 
as well as the variance of the front position
\begin{equation}
\sigma^2(Y)=\langle \rho_\text{s}^2(Y)\rangle 
-\langle \rho_\text{s}(Y)\rangle^2.
\end{equation}

\begin{figure}
\begin{center}
\epsfig{file=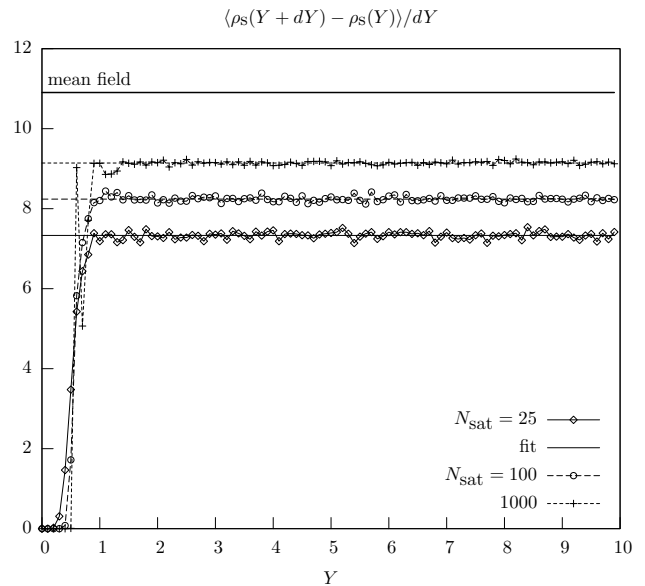,width=0.5\textwidth}
\end{center}
\caption{\label{plot_vfip}
Instantaneous velocity of the front in the FIP model, for $N_{\rm sat}=25,100,1000$
as a function of the rapidity.
The mean-field value of the velocity $\chi^\prime(\gamma_c)/\thelog B$
is also shown in full line.
}
\end{figure}

\begin{figure}
\begin{center}
\epsfig{file=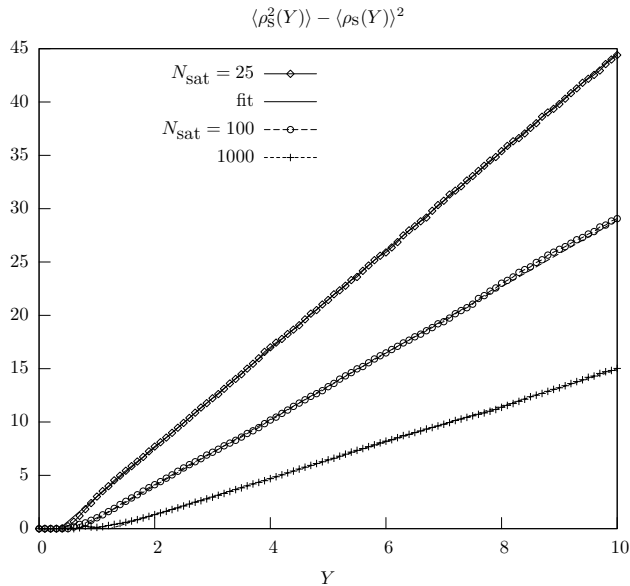,width=0.5\textwidth}
\end{center}
\caption{\label{plot_sigma2fip}
Variance of the front position as a function of the rapidity
in the FIP model.
Three different values of $N_\text{sat}$ are displayed. A linear fit is 
indistinguishable from the numerical data.
}
\end{figure}

First, let us address the FIP model 
in which there is no impact-parameter dependence.
We plot the velocity of the travelling wave for different values
of $N_\text{sat}$ in Fig.~\ref{plot_vfip}, and the corresponding
variance in Fig.~\ref{plot_sigma2fip}, 
both as a function of the evolution variable $Y$.
We see that the velocity reaches quite quickly its asymptotic
regime: it becomes stable as soon as $Y>2$, for any value of $j$.
The same holds for the growth rate of the variance.
A fit in the range $2<Y<5$ can safely be extrapolated to $Y>5$.
(This remark will be useful in the case of the AIP model, where we
have to restrict the calculation to $Y<5$ for technical reasons.) 
The results of the fits are shown in Tab.~\ref{data_aip}, for several
values of $N_\text{sat}$.

\begin{table}
\begin{tabular}{c|ccl|ccl|rrr}
&
\multicolumn{3}{c}{$\frac{d\langle\rho_\text{s}(Y)\rangle}{dY}$} & 
\multicolumn{3}{c}{$\frac{d\sigma^2}{dY}$} & \multicolumn{3}{c}{\# events}\\
\backslashbox{$N_\text{sat}$}{$\kappa$}  & FIP &
$10^{-2}$ & $10^{-3}$ & FIP & $10^{-2}$ & $10^{-3}$ & 
FIP & $10^{-2}$ & $10^{-3}$ \\
\hline\hline
10 & 6.42 & 7.55 &7.66& 5.92 & 4.27 & 4.22 & $10^4$ & 10360 & 28365\\
15 & 6.87 & 7.93 &8.04& 5.17 & 3.86 & 3.86 & $10^4$ & 10401 & 14245\\
25 & 7.33 & 8.26 &8.38& 4.62 & 3.07 & 3.06 & $10^4$ & 68602 & 11932 \\
50 & 7.84 & 8.60 &8.72& 3.76 & 2.44 & 2.50 & $10^4$ & 35127 & 7102\\
$100\phantom{^*}$& 8.24 & 8.87 &9.02& 3.11 & 1.99 & 2.25 & $10^4$ & 15851 & 3735 \\
$100^*$& --- & --- &$9.03^*$& --- & --- & $2.03^*$& --- & --- & $10^{4{*}}$\\
$200\phantom{^*}$& 8.57 & 9.11 & --- & 2.57 & 1.73 & ---   & $10^4$ &  8798 & {---} \\
\hline
\end{tabular}
\caption{\label{data_aip}
Velocity and rate of growth of the dispersion of travelling waves
in the FIP and AIP model, for different values of $N_\text{sat}$ and $\kappa$ 
(in the case of the AIP model). For $N_\text{sat}=100$ and $\kappa=10^{-3}$, 
we have also
performed a calculation up to $Y=4$ only (data denoted by a star) in order to
be able to collect enough statistics.}
\end{table}

\begin{figure}
\begin{center}
\epsfig{file=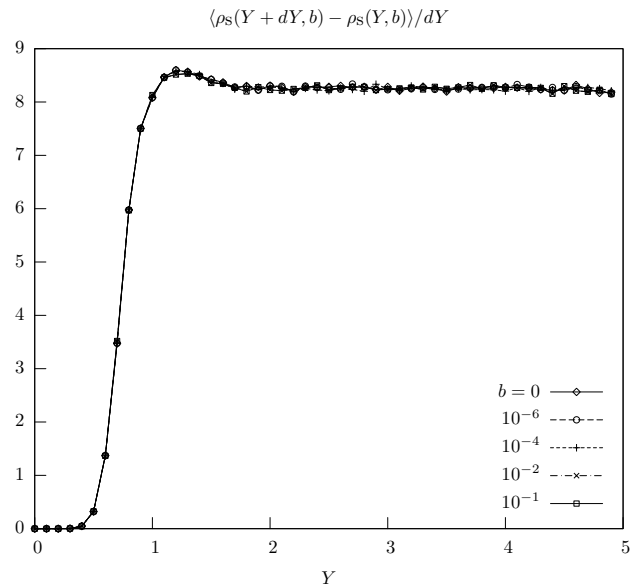,width=0.5\textwidth}
\end{center}
\caption{\label{plot_velocity_b}
Instantaneous velocity of the front in the AIP model, for $N_{\rm sat}=25$
and at different impact parameters. The curves are all almost perfectly
superimposed.
}
\end{figure}

\begin{figure}
\begin{center}
\epsfig{file=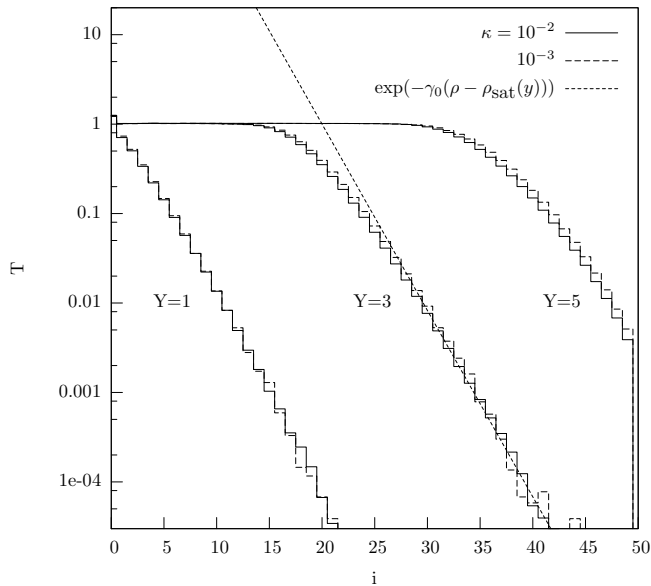,width=0.5\textwidth}
\end{center}
\caption{\label{plot_front2}
Average of the event-by-event amplitude 
in the AIP model
as a function of the dipole 
size at zero
impact parameter, for three different rapidities.
Two values of the cutoff $\kappa$ are compared and $N_{\rm sat}$ is set to 25.
}
\end{figure}

\begin{figure}
\begin{center}
\epsfig{file=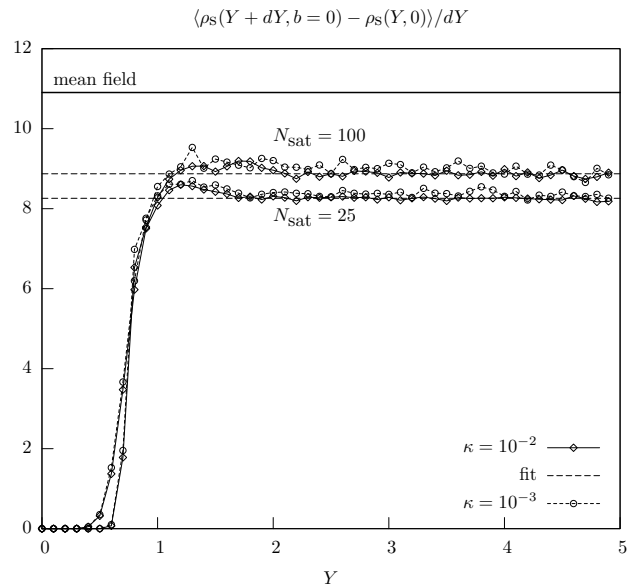,width=0.5\textwidth}
\end{center}
\caption{\label{plot_velocity}
Instantaneous velocity of the front in the AIP model, 
for $N_{\text{sat}}=25$ and $N_{\text{sat}}=100$, 
and
different values of the cutoff $\kappa$.
The impact parameter $b$ is set to $0$.
The mean-field result is also represented.
}
\end{figure}

\begin{figure}
\begin{center}
\epsfig{file=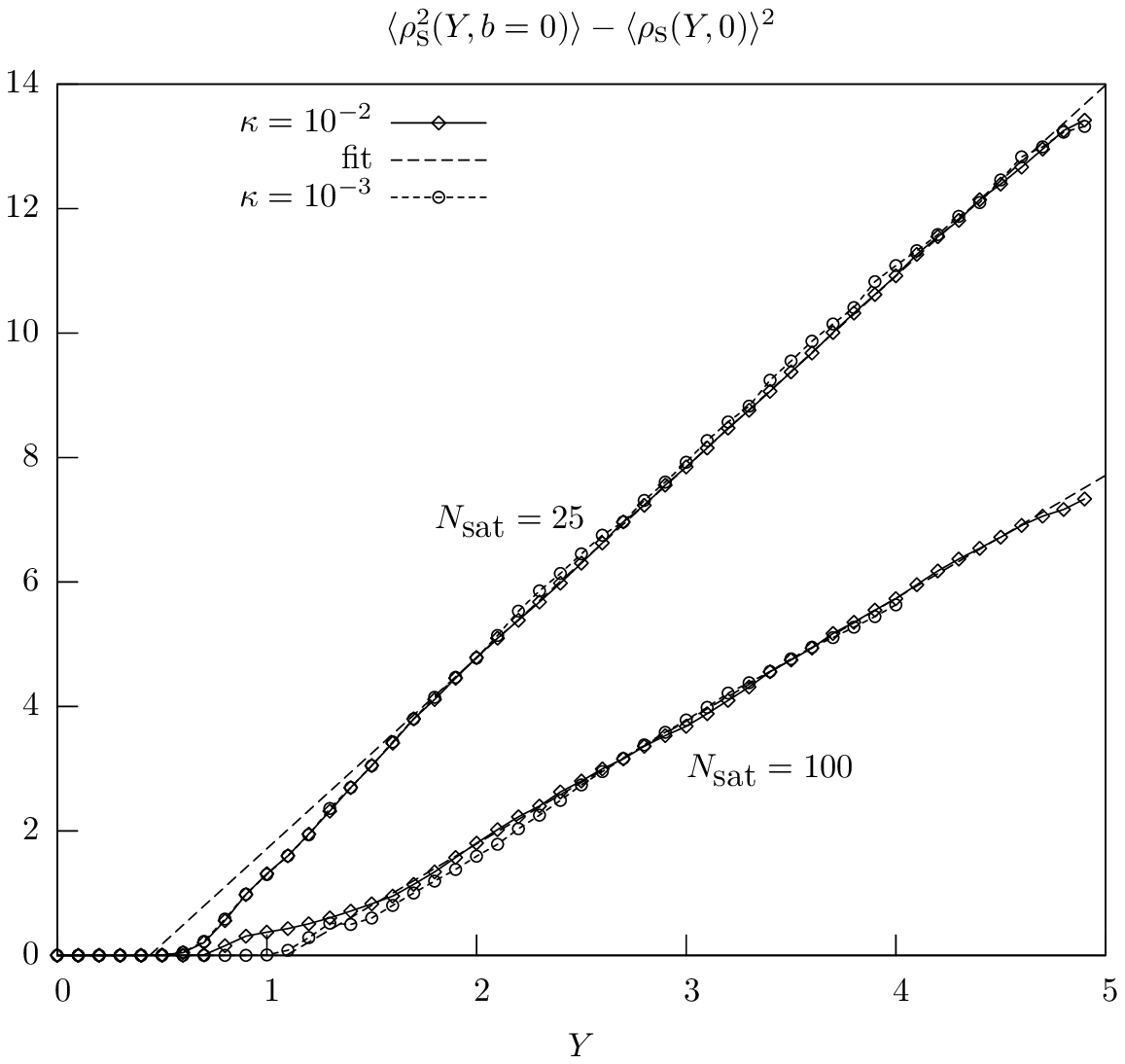,width=0.5\textwidth}
\end{center}
\caption{\label{plot_sigma2}
The same for the variance of the position of the front 
at central impact parameter. Note that for $N_{\text{sat}}=100$ and $\kappa=10^{-3}$,
we have data only up to $Y=4$.
}
\end{figure}

Now we repeat the same calculation in the AIP model, for $Y<5$. 
The choice of the upper limit on $Y$ is dictated by numerical limitations: 
if $Y$ gets larger, then the travelling waves are likely to explore too small
dipole sizes which cannot be resolved numerically for lack of accuracy 
(see Sec.~\ref{sec3A}).
We first check that the front velocity is independent of the 
impact parameter that is considered, Fig.~\ref{plot_velocity_b}.
As this is indeed the case, we are free to focus on central impact parameters.
Figure~\ref{plot_front2} shows travelling waves averaged over many events
at central impact parameter, for different values of the cutoff $\kappa$. 
(Note that the shape of the front in its forward part is consistent with 
the theoretical expectation in Eq.~(\ref{shape_front})).
Figures~\ref{plot_velocity} and~\ref{plot_sigma2} show the average of the
instantaneous velocity and of the dispersion of the front position
as a function of the rapidity.
Again, a stationary velocity is reached after 1 to 2 units of rapidity.
As soon as this regime is reached, the dispersion starts to
scale linearly in $Y$, as expected. The diffusion coefficient
$D=d\sigma^2/dY$, which is the slope of this line, quantifies the amount
of fluctuations.
The values of the velocity and of the diffusion coefficient are
reported in Tab.~\ref{data_aip}, also for different 
values of $N_\text{sat}$ (and $\kappa$).
The numbers have been extracted from a fit between $Y=2$ and 5 for the
velocity, and between $Y=2$ and $4.5$ for the diffusion
coefficient. We have limited the fit range for the
latter\footnote{Since the velocity instead depends on what happens close to
  the saturation scale, the fit can safely be extended to $Y=5$.} to
avoid edge effects related to the fact that the far tail of the front
sometimes reaches our cut-off on dipole sizes (see {\em e.g.}
Fig. \ref{plot_front2}), leading to the small turnover for the
dispersion observed in Fig. \ref{plot_sigma2}.

\begin{figure}
\begin{center}
  \includegraphics[angle=0,width=0.4\textwidth]{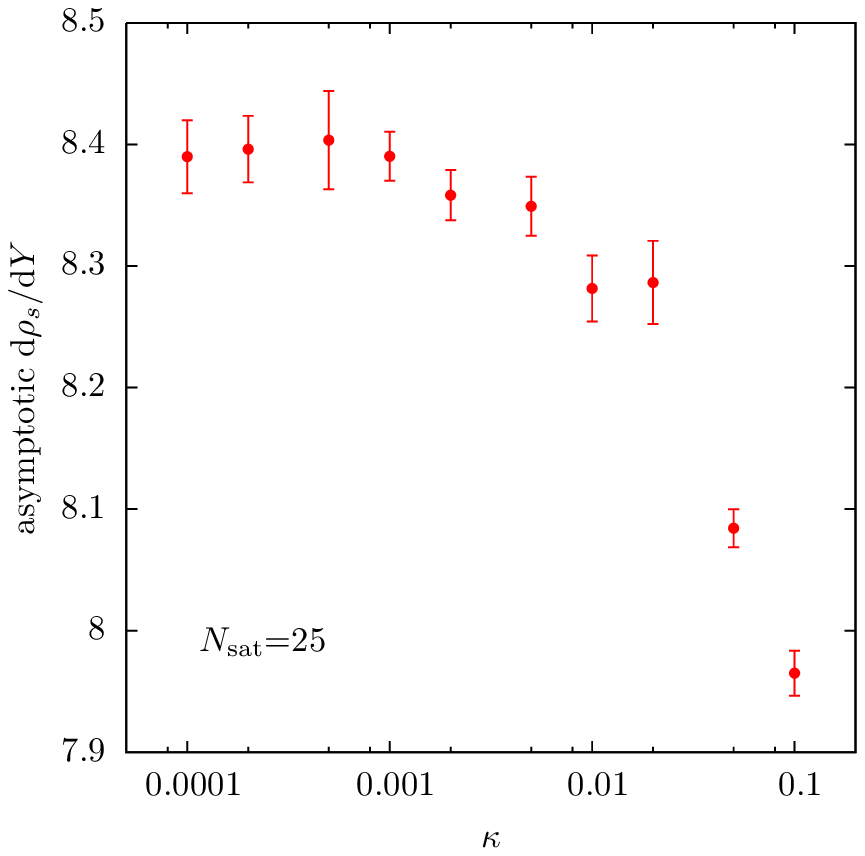}
\end{center}
\caption{\label{fig:kappa_dep}
$\kappa$ dependence of the asymptotic front velocity for $N_{\text{sat}}=25$.
}
\end{figure}

From Figs.~\ref{plot_velocity} and~\ref{plot_sigma2}, we already see
that there is no strong dependence on the value of $\kappa$. To
further study this dependence, we have computed the asymptotic
velocity for different values of $\kappa$ and $N_{\text{sat}}=25$. One
sees in Fig.~\ref{fig:kappa_dep} a saturation for $\kappa \lesssim 10^{-3}$ 
and a very small dependence on $\kappa$ up to $\sim 10^{-2}$,
confirming that the cutoff $\kappa$ does not strongly affect the
observables.

There is an important difference
between AIP and FIP, where similar results were expected
if the hypothesis of local evolution in impact parameter were literally correct.
Instead, 
comparing Figs.~\ref{plot_vfip} and \ref{plot_sigma2fip} with
Figs.~\ref{plot_velocity} and \ref{plot_sigma2},
we observe that the FIP model has more fluctuations than the
AIP model, as if the effective number of particles were lower, or
if some averaging were effectively performed in the AIP evolution.
In order to enforce the matching of the velocities and diffusion coefficients
of the two models, we should take a value of $N_\text{sat}$
of about 100 for the FIP model, {\em i.e.} about 4 times that of the
AIP model (with $N_{\text{sat}}=25$).
We will try and investigate more systematically this discrepancy below.
First, let us examine the correlations between travelling waves
at different impact parameters.


\subsection{Correlations in impact parameter}

\begin{figure}
\begin{center}
\epsfig{file=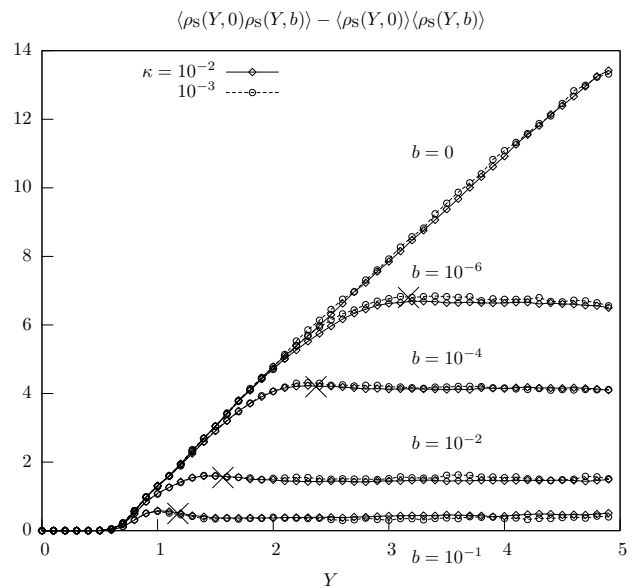,width=0.5\textwidth}
\end{center}
\caption{\label{plot_cor}
Correlations between the front position at different
impact parameters as a function of the rapidity for $N_\text{sat}=25$.
The crosses denote the rapidity $Y$ at which the (inverse of the) 
saturation scale 
$B^{-\rho_\text{s}(Y)}$ 
coincides with the distance in impact parameter space
between the probed points.}
\end{figure}

Fig.~\ref{plot_cor} represents the correlations between the
positions of the wave fronts at different impact parameters in the AIP model, defined
as
\begin{equation}
\langle \rho_\text{s}(Y,b_1)\rho_\text{s}(Y,b_2)\rangle-
\langle \rho_\text{s}(Y,b_1)\rangle\langle\rho_\text{s}(Y,b_2)\rangle.
\end{equation}
Once again, the different choices of $\kappa$ lead to very similar
results.

We also see very clearly the successive decouplings of the different
impact parameters, from the most distant to the closest one,
as rapidity increases. Indeed, the correlation functions flatten after some
given rapidity depending on the difference in the probed impact parameters, which
means that the evolutions decouple.
This decoupling is expected as soon as the travelling wave front
reaches dipole sizes which are smaller than the distance
between the probed impact parameters, {\em i.e.} at $Y$ such that
$|b_2-b_1|\approx 1/Q_s(Y) = B^{-\rho_\text{s}(Y)}$.
From the data for $\rho_\text{s}(Y)$, we can estimate quantitatively
the values of the rapidities at which the travelling waves decouple between the
different impact parameters. (It is enough to invert the above formula for the
relevant values of $b_2-b_1$).
These rapidities are denoted by a cross in Fig.~\ref{plot_cor} for the
considered impact parameter differences.
Our numerical results for the correlations are nicely consistent 
with this estimate, since the correlations start 
to saturate to a constant value precisely on the right of each such cross.


\subsection{$N_\text{sat}$-dependence and difference between AIP and FIP}

\begin{figure*}
\epsfig{file=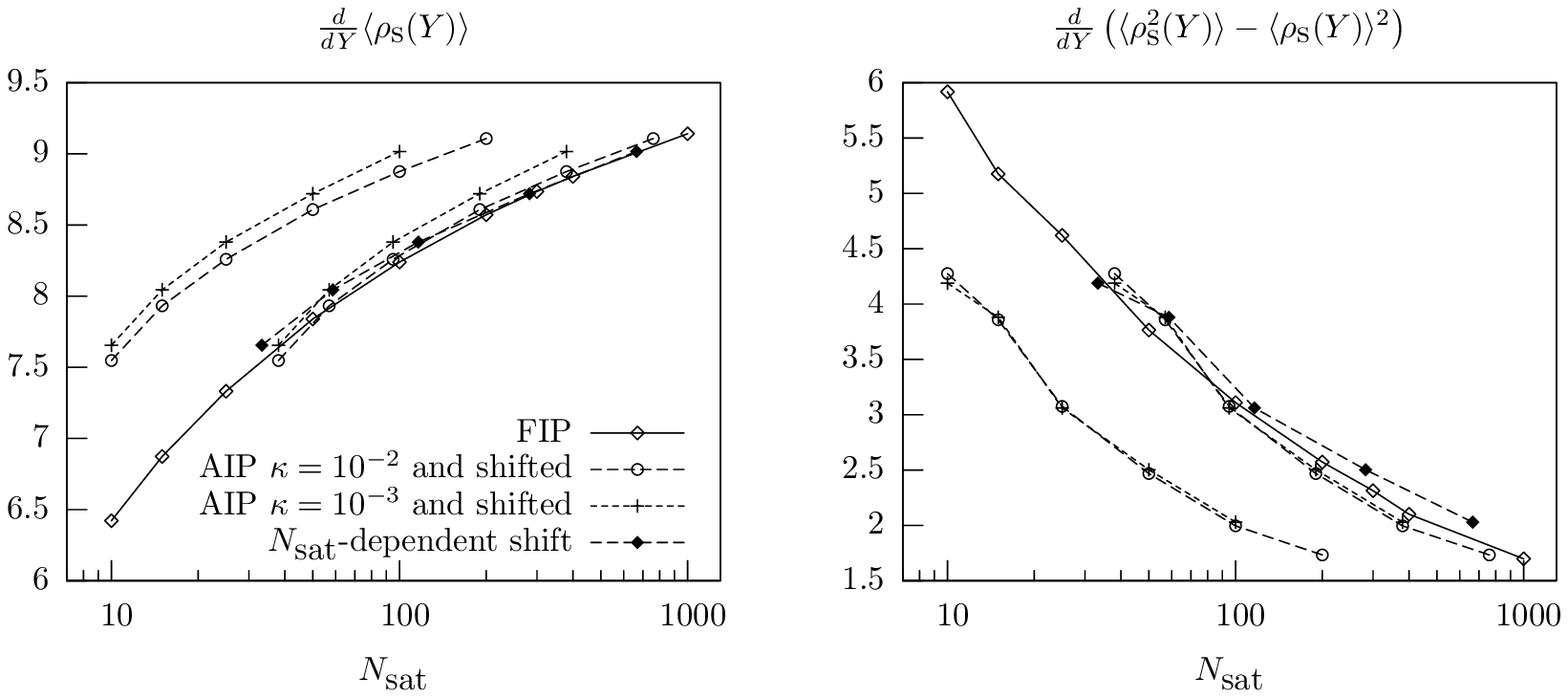,width=1\textwidth}
\caption{\label{plot_compare}
Steady-regime velocities (left) and diffusion coefficients (right) of the
wave front in the FIP model (full lines) and at central impact parameter
in the AIP model (dashed and dotted lines).
The two leftmost curves in each plot
correspond to the actual results of the calculation;
The two rightmost ones are the same but shifted. The shift is a scaling factor
of $N_\text{sat}=1/\alpha_s^2$ equal to 3.8.
The curves with filled black squared correspond to the AIP model
shifted by $\frac{1}{\gamma_c}\log N_\text{sat}$.
For $N_{\text{sat}}=100$ and $\kappa=10^{-3}$, we have performed a fit up to $Y=4$ only 
(see Tab.~\ref{data_aip}).
}
\end{figure*}

We have observed that the travelling wave fronts in the AIP model at
a given impact parameter look very much 
like those appearing in the FIP model, but the parameter
$N_\text{sat}$ should be changed by some factor to allow for a matching of their
characteristics.
In this section, we investigate more quantitatively this point.

To this aim, we compute the front velocity $V$ and diffusion coefficient
$D=d\sigma^2/dY$ in the AIP model
at central impact parameter as a function of $N_\text{sat}$.
We use different values of the cutoff $\kappa$ 
(in practice $\kappa=10^{-2}$ and $10^{-3}$).
A similar calculation can be performed
for the FIP model (in this case, there is no cutoff 
$\kappa$ to be considered).
The results 
are displayed in Fig.~\ref{plot_compare}.

We notice that the shapes of these curves look similar,
in the FIP and AIP model for $V$ as well as for $D$, 
except maybe for
very small values of $N_\text{sat}$ (10 and 15).
However, we may superpose the different curves only at the price
of performing a rescaling of $N_\text{sat}$ (which corresponds to a shift
of the curves on the logarithmic scale chosen in the plot 
of Fig.~\ref{plot_compare}).
Empirically, we find that the AIP model is equivalent to the FIP model
at each impact parameter if one sets
$N_\text{sat}$ for the latter about 3.8 times larger than for the former.

We may propose an interpretation of the observed discrepancy.
In the course of the evolution, travelling waves at different impact
parameters communicate by exchanging dipoles because dipole splitting
is not strictly local in coordinate space: starting from any given
step, the chain of subsequent splittings extends over an area that is
significantly larger than the area set by the initial dipole.
In this way, the waves may equalise their velocities,
and this leads to an extra dynamical 
averaging with respect to FIP models, at any impact parameter, continuously
at any step in the evolution.
This effective averaging results in a reduction of the 
fluctuations in the positions
of the waves, and thus to an increase of the front velocity, which
becomes closer to the mean-field velocity, and in particular different from
the expected velocities in the corresponding 
one-dimensional travelling wave model
(FIP model).

As we have observed in Fig.~\ref{plot_compare},
the scaling factor between $N_\text{sat}$ and the
effective maximum number of dipoles in the AIP model seems to be independent of
$N_\text{sat}$.
However, we also see that a factor $\log(N_\text{sat})/\gamma_c$ (the
length of the front) between AIP and FIP is also consistent with our
numerical observation. Unfortunately, we have not been able to find a
satisfactory analytic explanation for this factor.

In any case, since we find that the discrepancy amounts to a simple
rescaling of $N_\text{sat}$ by a constant or a slowly varying
factor (at least in the range of the values of
$N_\text{sat}$ that we were able to explore), it is likely that
it is a subleading effect, and that the
AIP and FIP models would agree asymptotically for large $N_\text{sat}$.
Hence we do not think that this slight mismatch between the AIP and
the FIP models spoils in any way the one-dimensional reaction-diffusion
picture of high-energy scattering.


\section{Conclusion}

One condition for high-energy evolution in QCD
to be in the universality class of a one-dimensional 
reaction-diffusion processes is that
the partonic evolution remains local in impact-parameter space.
In this paper, with the help of a toy model (called ``AIP'') engineered 
to incorporate the essential features of QCD and to
be manageable numerically,
we have checked that this is indeed the case.
We have measured the properties of travelling
waves at given impact parameters, and showed that the correlations
between waves at different points in coordinate space 
vanish when the rapidity gets large enough,
as expected for a local evolution.

Furthermore, we have compared the properties of the
travelling waves that appear in this model with the features
of their counterparts in a model that has no impact parameter
dependence (called ``FIP''), 
but which is identical to the former in all other respects.
We have found that the results are similar, except that
the number of partons at saturation should be
larger by a factor of the order of 4 in the FIP model, suggesting that the latter
has more fluctuations. 
It is also plausible that this factor has a slow $N_\text{sat}$-dependence
and may vary like $\frac{1}{\gamma_c}\log N_\text{sat}$.
At this stage, our numerical data do not enable us to check 
whether one of these guesses is literally correct.

We do not have a final explanation for the discrepancy between the AIP and
FIP models yet. However, the fact that the factor 
by which $N_\text{sat}$ should be scaled for the matching to occur 
does not exhibit a strong $N_\text{sat}$-dependence
(remember that $N_\text{sat}\sim 1/\alpha_s^2$ in QCD)
leads us to think that it is a subleading effect in the 
large-$N_\text{sat}$ (small-$\alpha_s$) limit, and thus,
that the one-dimensional reaction-diffusion picture is indeed valid, up to
the relevant replacement of the parameter $N_\text{sat}$
by some effective number of particles relevant to the propagation
of the travelling waves, yet to be understood more precisely. 

Note that, though a full study with two transverse degrees of freedom
would be of great interest, we believe that our one-dimensional
picture grasps the important aspects of the problem and, based on
universal properties of the reaction-diffusion systems, we expect our
results to hold for full QCD.

Finally, we can think of several other improvements beyond our study.
We have considered only one saturation mechanism, which
consisted in merely vetoing parton splittings in phase space cells
already occupied by at least $N_\text{sat}$ particles.
We could try and check whether other saturation mechanisms 
(such as parton recombination)
would bring about more correlations in impact parameter.
It would also be desirable to achieve some analytical understanding, in particular
of the rescaling factor between the AIP and FIP model. To this aim, it may
be interesting to study the chains of successive dipole splittings in the AIP model,
in order to find out whether rare fluctuations spread in a wider-than-expected
area in impact parameter.


\acknowledgments

The work of S.M. is supported in part by the
Agence Nationale pour la Recherche (France),
contract ANR-06-JCJC-0084-02. G.S. is supported by Contract
No. DE-AC02-98CH10886 with the U.S. Department of Energy.
We thank the Galileo Galilei Institute for Theoretical
Physics for the hospitality and the INFN for partial support when
this work was being initiated.

\end{document}